\documentstyle[12pt]{article}
\def\ln{{\rm ln\!}}
\def\n{\underline n}
\def\~{\tilde}
\def\em{\epsilon_-}
\def\Tr{{\rm Tr}}
\def\det{{\rm Det}}
\def\const{{\rm Const}}
\def\aoaq{{a_1 \ldots a_q}}
\def\bobq{{b_1 \ldots b_q}}

\def\.{\dot}

\def\a{\begin{eqnarray}}
\def\b{\end{eqnarray}}
\def\0{\nonumber}
\def\ba{\begin{array}}
\def\ea{\end{array}}

\def\q{{\bar{\cal Q}}}
\def\m{{\bar{\cal M}}}

\def\al{{\alpha}}
\def\lm{{\lambda}}

\def\cm{{\cal M}}

\renewcommand{\theequation}{\thesection.\arabic{equation}}

\newlength{\extraspace}
\newlength{\extraspaces}
\newcounter{dummy}
\newcommand{\ai}{
\addtocounter{equation}{1}
\setcounter{dummy}{\value{equation}}
\setcounter{equation}{0}
\renewcommand{\theequation}{\thesection.\arabic{dummy}\alph{equation}}
\begin{eqnarray}
\addtolength{\abovedisplayskip}{\extraspaces}
\addtolength{\belowdisplayskip}{\extraspaces}
\addtolength{\abovedisplayshortskip}{\extraspace}
\addtolength{\belowdisplayshortskip}{\extraspace}}
\newcommand{\bj}{
\end{eqnarray}
\setcounter{equation}{\value{dummy}}
\renewcommand{\theequation}{\thesection.\arabic{equation}}}
\def\d{{\partial}}

\newcommand{\ddlm}[1]{{\partial \over \partial \lm_{#1}}}

\begin{document}
\begin{flushright}
SISSA-ISAS 68/95/EP\\
hep-th/yymmxxx
\end{flushright}
\vskip0.5cm
\centerline{\LARGE\bf Multi--matrix models:}
\vskip0.3cm
\centerline{\LARGE\bf integrability properties and topological content}
\vskip1cm
\centerline{\large  L.Bonora, F.Nesti}
\centerline{International School for Advanced Studies (SISSA/ISAS)}
\centerline{Via Beirut 2, 34014 Trieste, Italy, and}
\centerline{INFN, Sezione di Trieste.  }
\vskip1cm
\centerline{\large E.Vinteler}
\centerline{International School for Advanced Studies (SISSA/ISAS)}
\centerline{Via Beirut 2, 34014 Trieste, Italy}
\vskip5cm
\abstract{We analyze multi--matrix chain models. They can be considered as
multi--component Toda lattice hierarchies subject to suitable coupling
conditions. The extension of such models
to include extra discrete states requires a weak form of integrability.
The discrete states of the $q$--matrix model are organized in representations
of $sl_q$. We solve exactly the Gaussian--type models, of which we
compute several all-genus correlators. Among the latter models one can
classify also the discretized $c=1$ string theory, which we revisit using
Toda lattice hierarchy methods. Finally we analyze the topological field theory
content of the $2q$--matrix models: we define primary fields (which are
$\infty^q$), metrics and structure constants and prove that they satisfy
the axioms of topological field theories. We outline a possible method
to extract interesting topological field theories with a finite number of
primaries.}

\vfill\eject

\section{Introduction}
\setcounter{equation}{0}
\setcounter{subsection}{0}

In this paper we intend to analyze matrix models made of $q$ Hermitean
$N\times N$ matrices with bilinear couplings between different matrices.
Unless otherwise specified, by this we mean an open chain of  $q$ matrices,
each linearly interacting with the nearest neighbours. These models have been
already introduced and partially analyzed in \cite{BX1} (for other approaches
to multi--matrix models, see \cite{Douglas},\cite{FK},\cite{DB},\cite{kostov},
\cite{AS},\cite{kharchev}).
The reasons to go beyond two--matrix models are diverse. The extended
two--matrix model provides
a useful representation of $c=1$ string theory at the self--dual point,
\cite{BX2}; in particular it naturally incorporates the so--called discrete
states, which appear to be organized in $sl_2$ multiplets. We find it
natural to ask ourselves whether such a construction can be generalized.
The answer is affirmative: in the extended $q$--matrix model we do find
discrete states organized according to representations of $sl_{q}$.
More recently it has been shown, \cite{BX3}, that $c=1$ string theory at the
self--dual point,
i.e. the two--matrix model, is a huge topological field theory in which
we can distinguish primaries, puncture operators and descendants.
As we shall see, this holds for $2q$ matrix model too, although with
new features (for example, the number of primaries is $\infty^n$).

On a more speculative ground one may remark that
two--matrix models lead via hamiltonian reduction to reduced models
characterized by classical
hierarchies \cite{BX3},\cite{AK} which can be interpreted in terms of
topological field theories coupled to topological gravity; in turn the latter
can be put in correspondence with string vacua. The two--matrix model
analysis suggests that, if we want to reach more interesting string or
W--string vacua, we have to shift to matrix models with several matrices.
Although we do not go as far as  proving this, nevertheless many elements
we find seem to support such a conjecture.

Finally, to end the list of the reasons of interest on a  more formal ground,
we recall that the integrable hierarchy characterizing two--matrix models
is the discrete Toda hierarchy, while the discrete integrable hierarchy
characterizing multi--matrix models is a generalization of the latter.
As we have already pointed out, we can extend these models by introducing
additional (extra) states and couplings. While the extended two--matrix model
does not present any essentially new features, the $q$--matrix models with
$q>2$ do. In fact we can have in general only a {\it weak} form of
integrability of the extra flows (as opposed to the {\it strong} integrability
of the ordinary cases). This form of integrability is nevertheless sufficient
for all our purposes.

The paper is organized as follows. In section 2 we review, mostly from
\cite{BX1}, the main results concerning multi--matrix models and derive
the flows of the extended $q$--matrix models. In section 3 we solve the
coupling condition of Gaussian $q$--matrix models. In section 4 we introduce
the
discrete states and discuss their group properties. We then compute several
examples of correlators in $2q$--matrix models. Section 5 is devoted to the
topological field theory properties alluded to above. In section 6
we introduce a few simple examples of non--Gaussian matrix models. Finally
section 7 is devoted to an analysis of the discretized 1D string. The latter
can in fact be envisaged as a chain matrix model with bilinear couplings.
It is interesting to rederive known properties of $c=1$ string in our
formalism.
Finally two Appendices are devoted to the $W$--constraints in $q$--matrix
models
and to an explicit computation, respectively.

\section{Multi--matrix models: general introduction}
\setcounter{equation}{0}
\setcounter{subsection}{0}

We review here some general results concerning $q$--matrix models, \cite{BX1}.
The partition function of the $q$--matrix model is given by
\a
Z_N(t,c)=\int dM_1dM_2\ldots dM_qe^{TrU}\label{Z}
\b
where $M_1,\ldots,M_q$ are Hermitian $N\times N$ matrices and
\a
U=\sum_{\al=1}^qV_{\al}
+\sum_{\al=1}^{q-1}c_{\al,\al+1}M_{\al}M_{\al+1}\0
\b
with potentials
\a
V_{\al}=\sum_{r=1}^{p_{\al}}\bar t_{\al,r}M_{\al}^r\,\qquad \al=1,2\ldots,q
\label{V}
\b
The $p_{\al}$'s are finite positive integers.

We denote by
${\cal M}_{p_1,p_2, \ldots, p_q}$ the corresponding  $q$--matrix model.
It has become moreover customary to associate to the generic
$q$--matrix model (\ref{Z}) the Dynkin diagram $A_q$. Occasionally we will
stick to this convention and speak abou nodes and links.

We are interested in computing correlation functions (CF's)
of the operators
\a
\tau_{\al, k}=tr M_\al^k\0
\b
and possibly of other composite operators (see below).
For this reason we complete the above model by replacing
(\ref{V}) with the more general potentials
\a
V_\al = \sum_{r=1}^\infty t_{\al,r} M_\al^r, \qquad \al =1,\ldots q\label{Vgen}
\b
where $t_{\al , r} \equiv \bar t_{\al,r}$ for $r\leq p_\al$.

In other words we have embedded the original couplings $\bar t_{\al,r}$ into
infinite sets of couplings. Therefore we have two types of couplings.
The first type consists of those couplings (the barred ones)
that define the model: they
represent the true {\it dynamical} parameters of the theory; they
are kept non-vanishing throughout the calculations.
The second type encompasses the remaining couplings, which
are introduced only for computational purposes.
In terms of ordinary field theory the former are analogous to the
interaction couplings, while the
latter correspond to external sources (coupled to composite operators).
Any CF is
obtained by differentiating $\ln Z_N$ with respect to the couplings associated
to the operators that appear in the correlator and then setting to zero
(only) the external couplings.

{}From now on we will not
make any formal distinction between interacting and external couplings.
Case by case we will specify which are the interaction couplings and which
are the external ones. Finally, it is sometime convenient to consider
$N$ on the same footing as the couplings and to set $t_{\al,0}\equiv N$.

The most popular procedure to calculate the partition function consists of
three steps \cite{BIZ},\cite{IZ2},\cite{M}:

\noindent
$(i)$. One integrates out the angular parts such that only the
integrations over the eigenvalues are left,
\a
Z_N(t,c)={\rm const}\int\prod_{\al=1}^q\prod_{i=1}^N d\lm_{\al,i}
\Delta(\lm_1)e^{{U}}\Delta(\lm_q),\label{Zint}
\b
where
\a
{U}=\sum_{\al=1}^q\sum_{i=1}^NV_\al(\lm_{\al,i})+
\sum_{\al=1}^{q-1}\sum_{i=1}^N c_{\al,\al+1}\lm_{\al,i}\lm_{\al+1,i},\label{U}
\b
and $\Delta(\lm_1)$ and $\Delta(\lm_q)$ are Vandermonde determinants.

\noindent
$(ii)$. One introduces the orthogonal polynomials
\a
\xi_n(\lambda_1)=\lambda_1^n+\hbox{lower powers},\qquad\qquad
\eta_n(\lambda_q)=\lambda_q^n+\hbox{lower powers}\0
\b
which satisfy the orthogonality relations
\a
\int  d\lambda_1\ldots d\lambda_q\xi_n(\lambda_1)
e^{\mu}
\eta_m(\lambda_q)=h_n(t,c)\delta_{nm}\label{orth1}
\b
where
\a
\mu\equiv\sum_{\al=1}^{q}\sum_{r=1}^{\infty}t_{\al,r}\lambda_{\al}^r
+\sum_{\al=1}^{q-1}c_{\al,\al+1}\lambda_{\al}\lambda_{\al+1}.\label{mu}
\b

\noindent
$(iii)$. If one expands the Vandermonde determinants in terms of these
orthogonal polynomials and using the orthogonality relation (\ref{orth1}), one
can easily calculate the partition function
\a
Z_N(t,c)={\it const}~N!\prod_{i=0}^{N-1}h_i\label{parti1}
\b
Knowing the $h(c,t)$'s amounts to knowing the partition function, up to
an $N$-dependent constant. In turn
the information concerning the $h(c,t)$'s can be encoded in suitable
{\it flow equations}, subject to specific conditions,
{\it the coupling conditions}. Before we come to that, however, we recall
some necessary notations.

For any matrix $M$, we define the conjugate ${\cal M}$
\a
\cm=H^{-1} MH,\qquad H_{ij}=h_i\delta_{ij},\qquad \bar M_{ij} = M_{ji},
\qquad M_l(j)\equiv M_{j,j-l}.\0
\b
As usual we introduce the natural gradation
\a
deg[E_{ij}] = j -i, \qquad {\rm where} \qquad (E_{i,j})_{k,l}= \delta_{i,k}
\delta_{j,l}\0
\b
and, for any given matrix $M$, if all its non--zero elements
have degrees in the interval $[a,b]$, then we will simply
write: $M\in [a,b]$. Moreover $M_+$ will denote the upper triangular
part of $M$ (including the main diagonal), while $M_-=M-M_+$. We will write
\a
{\rm Tr} (M)= \sum_{i=0}^{N-1} M_{ii}\0
\b
The latter operation will be referred to as taking the {\rm finite trace}.

\vskip0.2cm

{\bf Coupling conditions.}

First we introduce the $Q$--type matrices
\a
\int\prod_{\al=1}^q d\lm_{\al}\xi_n(\lambda_1)
e^{\mu}\lm_{\al}
\eta_m(\lambda_q)\equiv Q_{nm}(\al)h_m=\q_{mn}(\al)h_n,\quad
\al=1,\ldots,q.\label{Qal}
\b
Among them, $Q(1),\q(q)$ are Jacobi matrices: their pure upper triangular
part is $I_+=\sum_i E_{i,i+1}$.
We will need two $P$--type matrices, defined by
\a
&&\int\prod_{\al=1}^q d\lm_{\al}\Bigl(\ddlm 1 \xi_n(\lambda_1)\Bigl)
e^{\mu}\eta_m(\lambda_q)\equiv P_{nm}(1)h_m\\
&&\int  d\lambda_1\ldots d\lambda_q\xi_n(\lambda_1)
e^{\mu}\Bigl(\ddlm q \eta_m(\lambda_q)\Bigl)\equiv P_{mn}(q)h_n
\b

The matrices (\ref{Qalpha}) we
introduced above are not completely independent. More precisely all
the $Q(\alpha)$'s can be expressed in terms of only one of them and
one matrix $P$.
Expressing the trivial fact that the integral of the total derivative of the
integrand in eq.(\ref{orth1}) with respect to $\lm_{\al},1\leq\al\leq q$
vanishes, we can easily derive the constraints or {\it coupling conditions}
\ai
&&P(1)+V_1'+c_{12}Q(2)=0,\label{coup1}\\
&&c_{\al-1,\al}Q(\al-1)+V'_{\al}+c_{\al,\al+1}Q(\al+1)=0,\qquad
2\leq \al\leq q-1,\label{coup2}\\
&&c_{q-1,q}Q(q-1)+V_q'+{\bar {\cal P}}(q)=0.\label{coup3}
\bj
where we use the notation
\a
V'_{\al}=\sum_{r=1}^{p_{\al}}rt_{\al,r}Q^{r-1}(\al),\qquad \al=1,2,\ldots, q\0
\b

These conditions explicitly show that the Jacobi matrices depend
on the choice of the potentials. In fact they completely determine
the degrees of the matrices $Q(\al)$. A simple calculation shows that
\a
Q(\al)\in[-m_{\al}, n_{\al}],\qquad \al=1,2,\ldots,q\0
\b
where
\a
&&m_1=(p_q-1)\ldots(p_3-1)(p_2-1) \0\\
&&m_{\al}=(p_q-1)(p_{q-1}-1)\ldots(p_{\al+1}-1),\qquad~~~~~2\leq\al\leq q-1\0\\
&&m_q=1 \0
\b
and
\a
&&n_1=1\0\\
&&n_{\al}=(p_{\al-1}-1)\ldots(p_2-1)(p_1-1),
\qquad\qquad~~2\leq\al\leq q-1\0\\
&&n_q=(p_{q-1}-1)\ldots(p_2-1)(p_1-1)\0
\b
Throughout the paper we will refer to the following coordinatization of
the Jacobi matrices
\a
Q(1)=I_++\sum_i \sum_{l=0}^{m_1} a_l(i)E_{i,i-l}, \qquad\qquad\qquad
\q(q)=I_++\sum_i \sum_{l=0}^{m_q} b_l(i)E_{i,i-l}\label{jacobi}
\b
and for the supplementary matrices
\a
Q(\al)=\sum_i\sum_{l=-n_{\al}}^{m_{\al}}T^{(\al)}_l(i)E_{i,i-l},
\quad2\leq\al\leq q-1\label{midjacobi}
\b

\vskip0.2cm
{\bf Flow equations}

The flow equations of the q--matrix model can be expressed by means of
the following hierarchies of  equations for the matrices $Q(\al)$.
 \ai
&&{\partial\over{\partial t_{\beta,k}}}Q(\al)=[Q^k_+(\beta),
{}~~Q(\al)],\qquad 1\leq \beta\leq\al\label{CC1}\\
&&{\partial\over{\partial t_{\beta,k}}}Q(\al)=[Q(\al),
{}~~Q^k_-(\beta)],\qquad \al\leq \beta\leq q\label{CC2}
\bj
These flows commute and define a multi--component Toda lattice hierarchy,
\cite{UT},\cite{AS}.

\vskip0.2cm
{\bf Reconstruction formulae}.

The coupling conditions and the flow equations allow us to calculate
the matrix elements of $Q(\al)$. From the latter we can reconstruct
the partition function as follows. We start from the following main formula
\a
\frac {\d}{\d t_{\al,r}} \ln Z_N(t,c)={\rm Tr}\Bigl(Q^r(\al)\Bigl),
\qquad1\leq\al\leq q\label{parti2}
\b
It is evident that, by means of the flow equations for $Q(\al)$, we can
express all the derivatives of $\ln Z_N$ with respect to the couplings
$t_{\al, k}$ (i.e. all the correlators) as finite traces of commutators
of the $Q(\al)$'s themselves.
In other words, knowing the $Q(\al)$'s, we can reconstruct the partition
function (up to a constant depending only on $N$). In particular we can get
\a
{\d^2\over{\d t_{1,1}\d t_{\al,r}}}
\ln Z_N(t,c)=\Bigl(Q^r(\al)\Bigl)_{N,N-1},\qquad1\leq\al\leq q\label{parti3}
\b
It was already noticed in \cite{BX1} that this equation leads to the
two-dimensional Toda lattice equation.

\subsection{Extended $q$--matrix models.}

It  is important to be able to compute the correlators not only of the
states considered above, but also of new states, the {\it extra states}.
To this end we enlarge the $q$--matrix model by introducing in the
potential $U$ new interaction terms, as follows. We change
\a
U\rightarrow {\hat U} =\sum_{i=1}^N \sum_{b_1,\ldots,b_q} g_{b_1,\ldots,b_q}
\lambda_{1,i}^{b_1}\ldots \lambda_{q,i}^{b_q}\label{Uenlarged}
\b
in (\ref{Z},\ref{U}), and, accordingly,
\a
\mu\rightarrow \hat\mu=\sum_{b_1,\ldots,b_q} g_{b_1,\ldots,b_q}
\lambda_{1}^{b_1}\ldots \lambda_{q}^{b_q}\label{muenlarged}
\b
in (\ref{mu}). Henceforth $a_i,b_i,c_i,...$ will denote non--negative indices.

We denote by $\chi_{b_1,\ldots, b_q}$ the state specified (classically)
by $\sum_{i=1}^N  \lambda_{1,i}^{b_1}\ldots \lambda_{q,i}^{b_q}$. It
is clear that when $b_i =0$ for all $i\neq \al$, this state reduces to
$\tau_{\al,b_\al}$, while the corresponding coupling $g$ boils down to
$t_{\al,b_\al}$. Moreover the previously introduced bilinear coupling
$c_{\al,\al+1}$ is nothing but the above $g$ when all the $b_i=0$ except
$b_\al=b_{\al+1}=1$.

All the couplings and states that do not appear in
the original model (\ref{U}) are called {\it extra}. Exactly as in the original
$q$--matrix model, we can introduce orthogonal monic polynomials
$\xi_n(\lambda_1)$ and $\eta_m(\lambda_q)$ and define the $Q(\al)$ matrices.
This is parallel to what happens in the the extended two-matrix model,
\cite{BX2}.

However, unlike the extended two--matrix model, in the extended $q$--matrix
model, we cannot in general define flow equations in  matrix form like
eqs.(\ref{CC1},\ref{CC2}). This is a remarkable difference between extended
two-- and $q$--matrix models (with $q>2$), and,
at first sight, seems to spoil integrability and any possibility of exact
calculation of the CF's. Fortunately this is not the case. What one has
to do is not to calculate the flows of the matrices $Q(\al)$, but the
multiple derivatives w.r.t. the couplings of $\ln Z_N$, i.e. the multiple
derivatives of $h_n$, and express them in terms of matrices $Q(\al)$.
One can verify that such 'weak flows` commute, and
thus integrability is preserved, although in a weak sense.

The procedure is as follows. We first introduce two series of functions,
\cite{BX1},
\a
\xi^{(\al)}_n(t,\lm_{\al})\equiv\int\prod_{\beta=1}^{\al-1}
d\lm_{\beta}\xi_n(\lambda_1)e^{\mu_{\al}^L}.\label{xial}
\b
and
\a
\eta^{(\al)}_n(t,\lm_{\al})\equiv\int\prod_{\beta=\al+1}^{q}
d\lm_{\beta}e^{\mu_{\al}^R}
\eta_m(\lambda_q).\label{etaal}
\b
where
\a
&&\mu_{\al}^L\equiv\sum_{\beta=1}^{\al-1}\sum_{k=1}^{\infty}t_{\beta,k}
\lambda_{\beta}^k
+\sum_{\beta=1}^{\al-1}c_{\beta,\beta+1}\lambda_{\beta}\lambda_{\beta+1}.\0\\
&&\mu_{\al}^R\equiv\sum_{\beta=\al+1}^{q}\sum_{k=1}^{\infty}t_{\beta,k}
\lambda_{\beta}^k+\sum_{\beta=\al}^{q-1}c_{\beta,\beta+1}\lambda_{\beta}
\lambda_{\beta+1}.\0
\b
Obviously we have
\a
\xi^{(1)}_n(t,\lm_{1})=\xi_n(\lambda_1),\qquad\qquad
\eta^{(q)}_n(t,\lm_{q})=\eta_m(\lambda_q).\0
\b
but for other values of $\alpha$ one sees immediately that $\xi^{(\al)}$
and $\eta^{(\al)}$ are not polynomials. But they satisfy the orthogonality
relations
\a
\int d\lm_{\al}\xi^{(\al)}_n(t,\lm_{\al})
e^{V_{\al}(\lm_{\al})}\eta^{(\al)}_m(t,\lm_{\al})
=\delta_{nm}h_n(t,c),\qquad  1\leq\al\leq q.\label{orth3}
\b
Eq.(\ref{Qal}) provides a definition of the $Q(\al)$ matrix in this basis
\a
\int d\lm_{\al}\xi^{(\al)}_n(t,\lm_{\al})
\lm_\al e^{V_{\al}(\lm_{\al})}\eta^{(\al)}_m(t,\lm_{\al})
=Q_{nm}(\al)h_m(t,c),\qquad \forall 1\leq\al\leq q.\label{Qal1}
\b
Therefore the spectral equations follow
\a
&&\lm_{\al}\xi^{(\al)}=Q(\al)\xi^{(\al)},\qquad 1\leq\al\leq q.\label{specal}\\
&&\lm_{\al}\eta^{(\al)}=\q(\al)\eta^{(\al)},\qquad 1\leq\al\leq q.
\label{specal'}
\b
where $\xi^{\al}$ and $\eta^\al$ represent the infinite vectors with components
$ \xi^{\al}_0,\xi^{\al}_1,\ldots$ and $\eta^\al_0, \eta^\al_1, \ldots$,
respectively.

With these bases at hand one differentiates $h_n$, i.e. (\ref{orth1}) for
$n=m$,
w.r.t the appropriate couplings and evaluate the results when the extra
couplings vanish. The result contains derivatives of  $\xi_n$ and $\eta_n$
w.r.t to the couplings, which in turn can be evaluated differentiating
(\ref{orth1})
with $n>m$ or $n<m$. Finally one can express the result in terms of elements
of the matrices $Q(\al)$, by making use of the above
defined bases $\xi_n^\al$ and $\eta_m^\al$. Inserting this into the
expressions of the correlators, i.e. into the derivatives of $\ln Z_N$ w.r.t.
the appropriate couplings, one can express the latter in terms of finite
traces of polynomials in the $Q(\al)$'s.

{}From now on, whenever it is not confusing, we use the simplified notation
$Q(\al)\equiv Q_\al$.

The 1--point CF is easily found to be given by
\a
<\chi_{a_1,\ldots, a_q}>= \Tr \Big(Q_1^{a_1}\cdots Q_q^{a_q}\Big)\label{1p}
\b
The derivation of the two point functions, by the above
procedure, is as follows

$$<\chi_\aoaq\chi_\bobq>= \sum_{n=0}^{N-1} {\d^2 \ln h_n\over
\d g_\aoaq\d g_\bobq}$$
\a {\d^2 h_n\over \d g_\aoaq\d g_\bobq} &=&
	  \int d\lm{\d\over\d g_\bobq} \xi_n\lm_1^{a_1}...\lm_q^{a_q}\eta_n+
	  \int d\lm \xi_n\lm_1^{a_1+b_1}...\lm_q^{a_q+b_q}\eta_n+\0\\
&&+\int d\lm \xi_n\lm_1^{b_1}...\lm_q^{b_q}{\d\over\d g_\bobq}\eta_n \0\b
Then, using
$${\d\over\d g_\bobq} \xi_n =
-\sum_{m=0}^{n-1}(Q_1^{b_1}...Q_q^{b_q})_{nm}\xi_m,\qquad
  {\d\over\d g_\bobq}\eta_n =-
  \sum_{m=0}^{n-1}\eta_m (Q_1^{b_1}...Q_q^{b_q})_{mn} {h_n\over h_m},$$
we obtain
\a
<\chi_{a_1,\ldots, a_q}\chi_{b_1,\ldots,b_q}>&=&\Tr\Big[
Q_1^{a_1+b_1}\cdots Q_q^{a_q+b_q}
- \Big(Q_1^{b_1}\cdots Q_q^{b_q}\Big)_-\Big(Q_1^{a_1}\cdots Q_q^{a_q}\Big)
\0\\
&&-\Big(Q_1^{a_1}\cdots Q_q^{a_q}\Big)
\Big(Q_1^{b_1}\cdots Q_q^{b_q}\Big)_+\Big]\label{2p}
\b
Along the same lines, we get
\a
&&<\chi_{a_1,\ldots, a_q}\chi_{b_1,\ldots,b_q}\chi_{c_1,\ldots,c_q}>=
\Tr\Bigg\{Q_1^{a_1+b_1+c_1}\ldots Q_q^{a_q+b_q+c_q}\label{3p}\\
&&-\Bigg[\Big(Q_1^{a_1}... Q_q^{a_q}\Big)_-
\Big(Q_1^{b_1+c_1}... Q_q^{b_q+c_q}\Big)+
\Big(Q_1^{b_1+c_1}... Q_q^{b_q+c_q}\Big)
\Big(Q_1^{a_1}... Q_q^{a_q}\Big)_+ + \ {\rm c.p.}\Bigg]\0\\
&&+\Bigg[\Big(Q_1^{a_1}... Q_q^{a_q}\Big)_-
\Big(Q_1^{b_1}... Q_q^{b_q}\Big)\Big(Q_1^{c_1}... Q_q^{c_q}\Big)_+
+ \ {\rm p.}\Bigg]\0\\
&&+2 \Big(Q_1^{a_1}... Q_q^{a_q}\Big)_+
\Big(Q_1^{b_1}... Q_q^{b_q}\Big)_+\Big(Q_1^{c_1}... Q_q^{c_q}\Big)_+\Bigg\}.\0
\b
where p. (c.p.) means permutations (cyclic permutations) of the sets
$\{{a_1,...,a_q}\}$, $\{{b_1,...,b_q}\}$ and $\{{c_1,...,c_q}\}$.
The RHS's of both (\ref{2p}) and (\ref{3p})
are symmetric under the exchange of the $\chi$ operators. This property
together
with the fact that the RHS's can be written down in terms of the $Q_\al$'s,
which are calculable, expresses what we call {\it weak integrability}.

\section{Coupling conditions in Gaussian models.}

\setcounter{equation}{0}
\setcounter{subsection}{0}

There are several methods to solve matrix models. One is based on
$W$--constraints (see Appendix A) and will be occasionally used also in this
paper. The most powerful however
consists of solving the coupling conditions to obtain explicit expressions of
the $Q(\al)$ matrices and, then, inserting these into the expressions of the
correlators. In this paper we will mostly consider $q$--matrix models in
which the interacting terms are at most quadratic. These models
can be solved in general and the solutions can be expressed by means of
very general formulas (see below), the reason being that the coupling
conditions reduce to a system of linear equations in the $Q(\al)$'s.
We would like however to point out that much more general (than Gaussian)
$q$--matrix models can in
principle be exactly solved. The only trouble is that, when the potential terms
are more than quadratic, the coupling conditions are non-linear equations
in the $Q(\al)$'s and we cannot find such compact formulas as in the
Gaussian models but we have to proceed case by case. We will see later
on examples of non-Gaussian models. For the time being let us
concentrate on the Gaussian ones. They are sufficient to reveal the
topological properties of the corresponding matrix models.

Let us first introduce a more convenient notation for Gaussian models.
The q-matrix models with quadratic potential have the partition function
of the form:
\a
Z=\int \prod_{\al=1}^q dM_\al \exp(\sum_{\al=1}^q (t_\al M_\al^2+u_\al M_\al)+
\sum_{\al=1}^{q-1}c_\al M_\al M_{\al+1})\label{GaussZ}
\b
We notice that the linear terms can be eliminated by suitable redefinitions
of the matrices $M_\al$. However it is often useful to keep them distinct (for
example to study the topological field theory properties). Therefore,
whenever this does not complicate the formulas too much, we will keep the
linear terms.

We will solve first the coupling conditions of 3--
and 4--matrix models, both for pedagogical reasons and in order to have
explicit
formulas of the simplest cases, and then the general case.

\subsection{The 3-matrix model}

The coupling conditions are :
\a
P_1&&+2t_1Q_1+u_1+c_1Q_2=0\0\\
&&2t_2Q_2+u_2 +c_1Q_1+c_2Q_3=0\\
\overline{{\cal P}_3}&&+2t_3Q_3+u_3+c_2Q_2=0\0
\label{3matrix}
\b
Eliminating the matrix $Q$ we get the following two-matrix model type
coupling conditions:
\a
P_1&&+2(t_1-{c_1^2\over 4t_2})Q_1+(u_1-{u_2c_1\over 2t_2})-
{c_1c_2\over 2t_2}Q_3=0\0\\
\overline{{\cal P}_3}&&+2(t_3-{c_2^2\over 4t_2})Q_3+
(u_3-{u_2c_2\over 2t_2})-{c_1c_2\over 2t_2}Q_1=0\0
\b
Solving the system we obtain the following form of the $Q$ matrices
(with reference to the coordinatization (\ref{jacobi},\ref{midjacobi})
\a
b_0&=&-{2t_2\over B}\left(c_1c_2u_1-2c_2t_1u_2+
(4t_1t_2-c_1^2)u_3\right)\0\\
T_0^{(2)}&=&{2t_2\over B}\left(c_1t_3u_1-2t_1t_3u_2+c_2t_1u_3\right)\0 \\
a_0&=&-{2t_2\over B}\left( (4t_2t_3-c_1^2)u_1-2c_1t_3u_2+
c_1c_2u_3\right) \0
\b
and
\a
&&b_1=-{n\over B}\left(2t_2(4t_1t_2-c_2^2)\right),\qquad
a_1=-{n\over B}\left(2t_2(4t_2t_3-c_1^2)\right),\0\\
&&T_1^{(2)}={4c_1t_2t_3n\over  B},\qquad T_{-1}^{(2)}=-{2t_1n\over c_1},
\qquad R_3=-{2c_1c_2t_2n\over B}\0
\b
where
\a
B=(4t_2 t_3-c_2^2)(4t_2t_1-c_1^2)-(c_1c_2)^2\0
\b
while all the other coordinates vanish.

So far we have used a basis $\xi_n$ corresponding to the first matrix
or first node and to a basis $\eta_n$ corresponding  to the third matrix
or node. One may wonder what happens if one switches from the node 1,3 to
the nodes 1,2. The coupling constraints are of course modified:
$\overline{{\cal P}_3}$ disappears from the third eq. (\ref{3matrix}) and in
the
second eq.(\ref{3matrix}) there appears $\overline{{\cal P}_2}$ .Eliminating
now
the matrix $Q_3$ we obtain the 2--matrix model coupling conditions:
\a
P_1&&+2t_1Q_1+u_1-c_1Q_2=0\0\\
\overline{{\cal P}_2}&&+2(t_2-{c_2^2\over 4t_3})Q_2+(u_2-{u_3c_2\over 2t_3})
-c_1Q_1=0\0
\b
Calculating the form of the $Q$ matrices we get the same result as above.
Hence,changing the basis does not modify the model.
This is the simplest example of a base independence property which
must of course hold for all multi--matrix models with generic potentials.

\subsubsection{The 4-matrix model}

For the 4-matrix model the coupling conditions are:
\a
P_1+2t_1Q_1+u_1 +c_1Q_2&=&0\0\\
2t_2Q_2+u_2  +c_1Q_1+c_2Q_3&=&0\0\\
2t_3Q_3+u_3 +c_4Q_4+c_2Q_2&=&0\\
\overline{{\cal P}_4}+2t_4Q_4+u_4 +c_3Q_2&=&0\0
\label{4matrix}
\b
Eliminating the $Q_2$ and $Q_3$ matrices we get
the following constraints:
\a
P_1&&+2(t_1-{c_1^2t_3\over 4t_2t_3-c_2^2})Q_1+
(u_1+c_1{2t_3u_2-u_3c_2\over 4t_2t_3-c_2^2}) -
{c_1c_2c_3\over 4t_2t_3-c_2^2}Q_4=0 \0\\
\overline{{\cal P}_4}&&+2(t_4-{c_3^2t_2\over 4t_2t_3-c_2^2})Q_4+
(u_4-c_3{2t_2u_3-u_2c_2\over 4t_2t_3-c_2^2}) -
{c_1c_2c_3\over 4t_2t_3-c_2^2}Q_1=0\0\label{4matrix1}
\b
For $Q$ matrices we obtain the following form:
\a
a_0={s_{1}\over A },&\qquad&T_0^{(2)}={s_{2} \over A(t_2 t_3-c_2^2)},\0\\
b_0={s_{4}\over A},&&T_0^{(3)}={s_{3} \over A(t_2 t_3-c_2^2)},
\b
where
\a
A=4t_1t_2t_3t_4-4c_1^2t_3t_4-4c_2^2t_1t_4-4c_3^2t_1t_2+(c_1 c_3)^2
\b
while we do not write down the explicit expressions for the $s_\al$'s;
they are linear functions
of $u_{\al}$, therefore when $u_\al=0$ the $Q(\al)$ matrices are traceless.
The other coordinates are:
\a
T_{-1}^{(2)}=-{2t_1n\over c_1},\qquad
T_{-1}^{(3)}=n{4t_1 t_2-c_1^2\over c_1c_2},\qquad
R_4={c_1c_2c_3n\over A}\0
\b
and
\a
a_1&=&{2n(c_3^2t_2+c_2^2t_4-4t_2t_3t_4)\over A},\0\\
T_1^{(2)}&=&{ c_1n (4 t_3 t_4-c_3^2)\over A},\qquad
T_1^{(3)}={2c_1c_2t_4n\over A}\0\\
b_1&=&{2n(c_1^2t_3+c_2^2t_1-4t_2t_3t_1)\over A}.\0
\b

\centerline{---------------------}

Let us consider the two previous models at the {\it cosmological
point}, i.e. when all the couplings are set to zero
except the bilinear ones (the $c_\al$'s). The reason of the name is that
in such a case the CF's essentially depend only on $N$, which is
interpreted as the renormalized cosmological constant (see section 5).

We see immediately that, while such a point is well-defined for the 4--matrix
model, it is singular for the 3--matrix model (in fact $A\neq0$ but $B=0$).
These two models reveal the difference between odd and even $q$--matrix
models. The cosmological point is well-defined only for even
$q$--matrix models.

\subsection{Gaussian q--matrix models}

Let us now concentrate on the most general case (\ref{GaussZ}).
In particular $\mu$ takes the form
\a
\mu=\mu(\lambda_1,\ldots, \lambda_q) = \sum_{\al =1}^q u_\al \lambda_\al
+ \sum_{\al =1}^q t_\al \lambda_\al^2 + \sum_{\al =1}^{q-1}
c_\al \lambda_\al \lambda_{\al +1}\label{potGauss1}
\b
The coupling conditions are
\ai
&&P(1) +u_1 + 2t_1 Q(1) + c_1 Q(2) =0\label{Gcoupl1}\\
&& u_\al + 2t_\al Q(\al) + c_\al Q(\l +1) + c_{\al -1}Q(\al -1) =0 ,
\quad \al = 2,\ldots,q-1\label{Gcoup2}\\
&&\bar {\cal P}(q) + u_q + 2 t_q Q(q) + c_{q-1} Q(q-1) =0\label{Gcoup3}
\bj
These coupling conditions imply that $Q(\al)$ has only three non--vanishing
diagonal lines, the main diagonal and the two adjacent lines.
Now let us simplify the coordinatization
of such matrix as follows
\a
Q(\al) = \epsilon_+(\al) + \epsilon_0 (\al)+ \epsilon_{-}(\al)\label{Qalpha}
\b
where
\a
\epsilon_-(\al) = \sum_n g_\al(n)E_{n,n-1},
 \qquad \epsilon_0(\al) = \sum_n s_\al(n) E_{n,n}, \qquad \epsilon_+(\al)
 = \sum_n h_\al(n)E_{n,n+1}\0
\b
with the understanding that $h_1(n) =1$ and $g_q(n) = R(n)$.
In terms of these coordinates the above coupling equations take
the form of the following
linear system
\ai
&&2t_1 +c_1 h_2(n) =0 \0\\
&& 2t_1 s_1(n) + c_1 s_2(n)+u_1 =0\label{Glinsys1}\\
&& n + 2t_1 g_1(n) + c_1 g_2(n) =0 \0\\
&&2t_\al h_\al(n) + c_\al h_{\al+1}(n) + c_{\al-1}h_{\al -1}(n)=0,
\qquad \al =2,\ldots,q-1\0\\
&&  2t_\al s_\al(n) + c_\al s_{\al+1}(n)
+ c_{\al-1}s_{\al-1}(n)+u_\al =0,\quad\al =2,\ldots,q-1\label{Glinsys2}\\
&& 2t_\al g_\al(n) + c_\al g_{\al +1}(n)
+ c_{\al-1}g_{\al-1}(n)=0, \qquad\al =2,\ldots,q-1\0\\
&& \frac{n+1}{R(n+1)} + 2 t_q h_q(n) + c_{q-1}h_{q-1}(n)=0\0\\
&& 2t_q s_q(n) + c_{q-1}s_{q-1}(n) =0\label{Glinsys3}\\
&&2t_q R(n) + c_{q-1}g_{q-1}(n) =0\0
\bj
The solution of this system is expressed in terms of the matrices $X_\al$
and $Y_\al$, defined as follows
\a
X_\al = \left(\ba{cccccc} 2t_1 & c_1 & 0 &\ldots& 0 & 0 \\
			 c_1 & 2t_2 & c_2 &\ldots &0&0\\
			  0 &c_2 & 2t_3 &\ldots & 0&0\\
			  \ldots&\ldots&\ldots&\ldots&\ldots&\ldots\\
			  0&0&0& \ldots& 2t_{\al-1}& c_{\al-1}\\
			  0&0&0& \ldots& c_{\al-1}& 2t_\al\ea\right)\label{Xal}
\b
and
\a
Y_\al = \left(\ba{cccccc} 2t_\al & c_\al & 0 &\ldots& 0 & 0 \\
			 c_\al & 2t_{\al+1} & c_{\al+1} &\ldots &0&0\\
			  0 &c_{\al+1} & 2t_{\al+2} &\ldots & 0&0\\
			  \ldots&\ldots&\ldots&\ldots&\ldots&\ldots\\
			  0&0&0& \ldots& 2t_{q-1}& c_{q-1}\\
			  0&0&0& \ldots& c_{q-1}& 2t_q\ea\right)\label{Yal}
\b
Of course $Y_1\equiv X_q$. One finds

\a
&&h_\al(n) = (-1)^\al (c_1c_2\ldots c_{\al-1})^{-1} \det X_{\al-1}\0\\
&& R(n) = (-1)^q n c_1 c_2 \ldots c_{q-1} \Big(\det X_q\Big)^{-1}
\label{sollinsys}\\
&&g_\al(n) = (-1)^\al n c_1c_2\ldots c_{\al-1} \frac{\det Y_{\al+1}}{\det
X_q}\0
\b
Moreover, if we denote by $S$ and $U$ the vectors $(s_1,s_2,\ldots,s_q)^t$
and $(u_1,\ldots, u_q)^t$, respectively, we have
\a
S = -X_q^{-1}U \label{S}
\b
As we have already remarked we can always without
loss of generality suppress the linear terms in $u_\al$ by constant shifts of
$M_\al$. In such a case $S=0$.

It is now easy to see that, at the cosmological point ($t_\al = u_\al=0$),
the solution (\ref{sollinsys}) is well defined when $q$ is even, while it
is singular when $q$ is odd -- in the latter case, for example, $\det X_q=0$.

\centerline{------------------------}

In the last part of this section we would like to dispel a seemingly obvious
objection to the very content of this paper.
Take the generic quadratic model of $q$ matrices with nearest neighbour
interactions
\a
U = \sum_{\al=1}^qt_\al M_\al^2
	       + \sum_{\al=1}^{q-1}c_{\al,\al +1} M_\al M_{\al+1}
        	      \equiv \sum_{\al,\beta}^q M_\al A_{\al\beta}M_\beta.
		      \label{genpotGauss}
\b
The $q\times q$ matrix $A$ is symmetric, and, for the theory of
central quadrics, it can be brought to a canonical diagonal form
with all ones or minus ones on the diagonal. The signature of A
is of course a characteristic of the potential.

Let us see the consequences of this simple remark as far as the matrix model is
concerned. The diagonalization of $A$ can be achieved by integrating
in the path integral over suitable linear combinations of the matrices $M_\al$,
instead of integrating simply over the $M_\al$'s. Of course this
gives rise to a Jacobian factor, which is however one if one
uses only shifts of the $M_\al$'s. In this way one brings $A$ to the
diagonal form
\a
A = {\rm Diag}(f_1,\ldots, f_q)\label{diag}
\b
but does not rescale its elements to unity. However this form is
sufficient for our subsequent discussion. The initial matrix model appears
at this point to be equivalent to the decoupled model with potential
\a
U' = \sum_\al f_\al M^2_\al. \0
\b
with partition function  $Z=\const(N)(f_1f_2...f_q)^{-N^2/2}$. We remark
however
that this procedure is of no help if one has  to compute
correlation functions of composite operators, in that it screws up
the definition of the states and renders the calculation of the
correlators practically impossible. The procedure followed in this paper,
i.e. the use of the generalized Toda lattice hierarchy, has precisely the
virtue that it allows the calculation of the exact correlators of significant
composite operators.

Finally let us remark that we can easily generalize the results of this
subsection to the cases when in the potential are present, beside the terms
of (\ref{genpotGauss}), also interactions
of the type $c_{\al,\beta} D_\al D_\beta$ where $D_\al= {\rm Diag~M_\al}$
and $\beta\neq \al-1,\al,\al+1$. In such cases
the method is the same as in the chain models, with the only difference that
the matrices $X_\al$ and $Y_\al$ will have, at the position $(\al,\beta)$,
additional non--vanishing entries $c_{\al\beta}$ if the latter are present
in the potential.

\section{Correlation function of discrete $sl_q$ states}
\setcounter{equation}{0}
\setcounter{subsection}{0}

In the previous section we have shown how to solve the coupling conditions
of a given model. In this section we show how to calculate various correlation
functions of composite operators (or discrete states). To start with let
us illustrate a basic property of the latter: in the $q$ matrix model they are
organized in finite dimensional representations of $sl_q$.

\subsection{$sl_q$ symmetry of the discrete states.}

We have shown in section 2.1 that we can enlarge the $q$--matrix model
by introducing in the potential terms of the form:
\a
g_{a_1,\ldots a_{q}}\prod_{\al=1}^{q}D_\al^{a_\al}, \qquad {\rm with} ~
D_\al={\rm Diag}(M_\al) \0
\b
We call {\it discrete states} the operators $\chi_{a_1,\ldots a_{q}}$ coupled
to $g_{a_1,\ldots a_{q}}$.
We introduce also $\chi_{0,\ldots 0}\equiv Q$ as the operator coupled to
$g_{0,\ldots 0}\equiv N$. Classically, $\chi_{a_1,\ldots a_{q}}$ is
represented by $\sum_{k=1}^N\lambda_{1,k}^{a_1}\ldots\lambda_{q,k}^{a_{q}} $.
These states carry a built--in  $sl_{q}$ structure.
To see this one has to consider the following generators
\a
H_i={1\over 2}\sum_{k=0}^N\left(\lm_{i,k} {\d \over \d \lm_{i,k}}-
 \lm_{i+1,k} {\d \over
{\d \lm_{i+1,k}}}\right),\qquad 1\leq i\leq q-1\0
\b
\a
E_{i,j}^+=\sum_{i=1}^N \lm_{i,k} {\d \over \d\lm_{j,k}},\qquad
E_{i,j}^- = \sum_{i=1}^N \lm_{j,k} {\d \over \d \lm_{i,k}}, \qquad
1\leq i <j\leq q\0
\b
$H_i$ form the Cartan subalgebra of $sl_q$, while $E_{i,j}^+$ and $E_{i,j}^-$
are, respectively, the raising and lowering operators of the Lie
algebra $sl_q$, corresponding to the roots:
\a
\al_{ij}=\varepsilon_i-\varepsilon_j,\quad i<j\0
\b
in the standard notation. The action on the states is as follows:
\a
H_i \chi_\aoaq = {1\over 2}(a_i-a_{i+1})\chi_\aoaq,
\qquad E^\pm_{i,j} \chi_{a_1,\ldots, a_i,\ldots ,a_j,\ldots, a_{q}} =
 \chi_{a_1,\ldots ,a_i\pm 1,\ldots ,a_j\mp 1,\ldots, a_{q}}\0
\b
Therefore the set $\{\chi_{a_1,\ldots a_{q}}=\sum_{i=1}^N
\lambda_{1,k}^{a_1}\ldots\lambda_{q,k}^{a_{q}} ,~~ \sum_{i=1}^qa_i=n\}$ form
an (unnormalized) representation of this algebra of dimension
$\left(\ba{c}n+q-1 \\ n \ea\right)$.

\centerline{-------------------}

Although everything we do here
can be repeated for $q$--matrix model with $q$ odd, we concentrate from now on
on the far more interesting case of even $q$. The main reason for this is
the well-definedness of the cosmological point when $q$ is even.
This will allow
us to give an unambiguous topological field theory interpretation of the
corresponding matrix models, while such an interpretation does not  seem to
be possible for odd $q$. Therefore, from now on, unless otherwise specified
we consider $2q$--matrix models.

\subsection{General properties  of correlators}

The correlation functions of the extended multi-matrix model are
in general defined by
\a
<\chi_{a_1^{(1)},\ldots a_{2q}^{(1)}}\ldots \chi_{a_1^{(n)},
\ldots a_{2q}^{(n)}}> =
{\partial\over \partial g_{a_1^{(1)},\ldots a_{2q}^{(1)}}}
 \ldots {\partial\over\partial g_{a_1^{(n)},\ldots a_{2q}^{(n)}}} \ln Z_N\0
\b

Our purpose in this section is to calculate the correlation functions in
two simple special cases: {\it the pure chain models}
where we set $g_{a_1,\ldots a_{2q}}=0$
except for $g_{0\ldots a_\al a_{\al+1}\ldots 0}\equiv c_\al$
and {\it the quadratic models} where we have also the following
nonzero coupling constants $g_{0\ldots a_\al=2 \ldots 0} \equiv  t_\al,
g_{0\ldots a_\al=1\ldots 0}\equiv u_\al$.
As a consequence the CF's will be functions of $c_\al,t_\al,u_\al$ and $N$.
The chain models were referred to above as the {\it cosmological point} of
the relevant $2q$--mm, while the quadratic models can be considered as
quadratic perturbations of the latter. This second terminology is related
to the topological field theory interpretation of section 5.

To see some general properties of the CF's, it is convenient to use
the $W$--constraints (see Appendix A).
We write down the $W$ constraints in terms of them and obtain a set of
(overdetermined) algebraic equations which in general one can solve
recursively.

The CF's, in the chain models, have the following symmetry property:
\a
&&<\chi_{a_1^{(1)},\ldots a_i^{(1)}\ldots a_j^{(1)}\ldots a_{2q}^{(1)}}
\ldots\chi_{a_1^{(n)},\ldots a_i^{(n)}\ldots a_j^{(n)}\ldots a_{2q}^{(n)}}>\0\\
&&~~~~~~~~~~~~~~~~~~~~~~~~~
= <\chi_{a_1^{(1)},\ldots a_j^{(1)}\ldots a_i^{(1)}\ldots a_{2q}^{(1)}}
\ldots\chi_{a_1^{(n)},\ldots a_j^{(n)}\ldots a_i^{(n)}\ldots a_{2q}^{(n)}}>\0
\b
This is due to the symmetry of the $W$ constraints
and to the invariance of the chain models under the exchange
$i\leftrightarrow j$.

In the chain models the CF's satisfy (charge conservation):
\a
\sum_{\al=1}^q[(a_{2\al-1}^{(1)}+\ldots a_{2\al-1}^{(n)})-
(a_{2\al}^{(1)}+\ldots a_{2\al}^{(n)})]
<\chi_{a_1^{(1)}\ldots a_{2q}^{(1)}}\ldots\chi_{a_1^{(n)}\ldots
a_{2q}^{(n)}}>=0
\label{lemma}
\b

To prove the last statement we rewrite the $W_0^{[1]}$ constraint as follows:
\a
&&\sum_{a_1\geq 1,a_2\geq 0\ldots a_{2q}\geq 0} a_1g_{a_1\ldots a_{2q}}
 <\chi_{a_1\ldots a_{2q}}> + {1\over 2}
N(N+1) =0\0\\
&&\sum_{a_1\geq ,\ldots a_\al\geq1\ldots a_{2q}\geq 0}
 a_\al g_{a_1\ldots a_{2q}} <\chi_{a_1\ldots a_{2q}}> =0,\qquad 2\leq \al\leq
 2q-1\0 \\
&&\sum_{a_1\geq 0,a_{2q-1}\geq 0 a_{2q}\geq 1} a_qg_{a_1\ldots a_{2q}}
 <\chi_{a_1\ldots a_{2q}}> + {1\over 2}
N(N+1) =0\0
\b
 We differentiate these equations w.r.t.
$g_{a_1^{(1)}\ldots a_{2q}^{(1)}}$, \ldots,
$g_{a_1^{(n)}\ldots a_{2q}^{(n)}}$ and set $g_{a_1\ldots a_{2q}}=0$
 except $g_{0\ldots a_{\al}a_{\al+1} 0}=c_\al$. One gets
\a
&&\sum_{k=1}^na_1^{(k)}
<\chi_{a_1^{(1)}\ldots a_{2q}^{(1)}}\ldots\chi_{a_1^{(n)}\ldots a_{2q}^{(n)}}>
+c_1<\chi_{110\ldots 0}
\chi_{a_1^{(1)}\ldots a_{2q}^{(1)}}\ldots\chi_{a_1^{(n)}\ldots a_{2q}^{(n)}}>
=0\0\\
&&\sum_{k=1}^na_\al^{(k)}
<\chi_{a_1^{(1)}\ldots a_{2q}^{(1)}}\ldots\chi_{a_1^{(n)}\ldots a_{2q}^{(n)}}>
+<(c_{\al-1}\chi_{0\ldots 1,a_\al=1\ldots 0}+\0\\
&&~~~~~~~~~~~c_{\al}\chi_{0\ldots a_\al=1,1\ldots 0})
\chi_{a_1^{(1)}\ldots a_{2q}^{(1)}}\ldots
\chi_{a_1^{(n)}\ldots a_{2q}^{(n)}}>=0, \qquad 2\leq \al\leq 2q-1\0\\
&&\sum_{k=1}^na_{2q}^{(k)}
<\chi_{a_1^{(1)}\ldots a_{2q}^{(1)}}\ldots\chi_{a_1^{(n)}\ldots a_{2q}^{(n)}}>
+c_{2q}<\chi_{0\ldots 011}
\chi_{a_1^{(1)}\ldots a_{2q}^{(1)}}\ldots\chi_{a_1^{(n)}\ldots a_{2q}^{(n)}}>
=0\0
\b
Subtracting the even equations from the odd ones we obtain the result.

The last property partially reflects the $sl_{2q}$ structure of the discrete
states as it means, at the cosmological point, the conservation of the
eigenvalue of $H=H_1+H_3+\cdots+H_{2q-1}$.

\centerline{---------------}

In the remaining part of this section we are going to compute exact
correlators, i.e. all--genus expressions, from which we can
extract the genus by genus expansion. To obtain such an expansion
it is of fundamental
importance that we can assign to each coupling a degree, denoted $[\cdot]$,
as follows
\a
\relax [g_{a_1,...,a_q}] = y +\sum_{i=1}^q y_ia_i, \qquad [N]=y\label{deg1}
\b
and for each quantity
such as free energies, correlators, etc., we can define a genus expansion,
each genus contribution having a definite degree,
\a
F=\ln Z_N ,\qquad F=\sum_{h=0}^\infty F_h,\qquad [F_h] = 2(1-h)y\label{deg2}
\b
In eqs.(\ref{deg1},\ref{deg2}), $y,y_i$ are arbitrary nonvanishing real
numbers.

\subsection{The 1-point correlation functions}

To find explicit expressions of the correlators it is more convenient to switch
from the $W$--constraints to the method based on the solution of the coupling
conditions.

\subsubsection{Pure chain models}

We specialize here to the case in which all the couplings, except
the bilinear ones, are turned off:
\a  t_{r,\al} = 0. \b
In this case the coupling conditions take the form:
\a
\label{Ccond}
P_1                +c_1 Q_2          &=&0,\\ \0
c_{\al-1}Q_{\al-1}+c_{\al}Q_{\al+1}&=&0 \ \ \ (\al=2,\ldots,  2q-1)\\
\bar{{\cal P}_{2q}}          +c_{2q}Q_{2q-1}       &=&0 \0
\b

This linear system is so simple that we do not need to rely on the formulas of
the previous section. We note that $Q_1$ has only the first
upper diagonal, and $P_1$, which represents a derivative,
has only the first lower one
$$
Q_1 = I_+ =\left( \ba{ccccc}
0&1& & & \\
 & &1& & \\
 & & &1& \\
 & & & &\ddots\ea\right)  \ \ \
P_1 = \em =\left( \ba{ccccc}
0& & & & \\
1& & & & \\
 &2& & & \\
 & &3& & \\
 & & &\ddots& \ea\right)
$$

The first coupling condition (\ref{Ccond}) gives now
$Q_2=-1/c_1\em$. Using the second for $\alpha=3$ one finds
then $Q_4$, and so on for all the even $Q$'s up to $Q_{2q}$.
For the odd ones the procedure is the same starting from $Q_1$.
\a
Q_{2k+1} = (-1)^k{c_1 \over c_2} \dots {c_{2k-1} \over c_{2k}} I_+\0\\
  Q_{2k} = (-1)^k{c_2 \over c_1} \dots {1 \over c_{2k-1}} \em
\label{Qsolution}\b
Now we come to the correlation functions which are expressed in terms
of the $Q$-matrices by means of the formula:
\a
<\chi_{a_1,...,a_{2q}}>=\Tr({Q_1}^{a_1} {Q_2}^{a_2} \dots Q_{2q}^{a_{2q}}).
\label{onepointQ}
\b

Due to the particular form of the $Q$'s it is immediate to verify
the conservation law (\ref{lemma}). In order
to have nonvanishing trace the number of $I_+$
and $I_-$ must be the same, i.e.
\a
a_1+a_3+\dots+a_{2q-1}-a_2-a_4-\dots-a_{2q}=0 \0
\b

\medskip

The result for the one point functions can be found by means of the
identities $[I_+,\em]=I_0$ and $I_+\em=\n$ ($I_0$ being the identity
matrix and $\underline n=diag(1,2,3,...)$):
\a
<\chi_{a_1,...,a_{2q}}>&=\sum_{n=x}^{N-1}
			& (n+a_1-a_2)(n+a_1-a_2+1)\dots(n+a_1-1)\0\\
		       &&    (n+a_1-a_2+a_3-a_4)\dots(n+a_1-a_2+a_3-1)\dots\0\\
		       &&    (n+a_1-a_2+\dots+a_{2q-1}-a_{2q})\dots(n+a_{2q}-1)
\b
where $x=-{\rm min}[a_1-a_2,a_1-a_2+a_3-a_4,\dots,a_1-a_2+...-a_{2q}=0]$.

\subsubsection{Quadratic models}

We write down some special 1-point correlation functions in the quadratic
models. The derivation can be found in Appendix B.

\a
<\tau_{\al,r}>=\mbox{Tr}(Q_\al^r)=\sum_{2l=0}^{r}\sum_{k=0}^{l}
{(-1)^k 2^{-k}r!\over (r-2l)!k!(l-k)!}\left(\ba{c}N+l-k\\l-k+1\ea \right)
(h_\al g_\al)^l s_\al^{r-2l}
\b
where $h_\al, g_\al$ are defined by eqs.(\ref{sollinsys}).

This is an all--genus expression. In order to extract the genus h contribution
follow the above described recipe. In particular for
the 1-point functions we have the following expansion
\a
<\tau_{\al,k}>=\sum_{h=0}^\infty <\tau_{\al,k}>_h N^{1+k-2h}\0
\b
The  only
dependence on $N$ comes from $\left(\ba{c} N+r \\ r=1 \ea \right)$ and  we
can expand it as follows
\a
\left(\ba{c}  N+r  \\  r+1  \ea  \right)=\sum_{h=0}^\infty  N^{1+k-2h}
b_{2h}(k)
\b
Using the last relation we can extract the genus $h$ contribution:
\a
<\tau_{\al,r}>_h =\mbox{Tr}(Q_{\al}^r) &=&\sum_{2l=0}^{r}\sum_{k=0}^{l}
{(-1)^k 2^{-k}r!b_{2h}(l-k+1)\over (r-2l)!k!(l-k)!}
N^{l-k+1-2h}\cdot\\
&&\cdot (h_\al g_\al)^l s_\al^{r-2l}\0
\b
with:
$$b_{2h}(n)=\sum_{1\leq r_1\ldots r_{2h}\leq n}r_1r_2\ldots r_{2h},\qquad
b_0(n)=1$$

\subsection{Two-point functions}
\subsubsection{Pure chain models}

For the 2--point functions we have to use eq.(\ref{2p}).
As an example we calculate the correlation functions of the form:
	  $$<\chi_{0...a_\al=r...0}\chi_{0...a_\beta=s...0}>
		  =<\tau_{\al,r}\tau_{\beta,s}>$$

The formula (\ref{2p}) becomes, in this case,

	$$<\chi_{0...a_\al=r...0}\chi_{0...a_\beta=s...0}>
        =\left\{ \ba{c}
        \Tr([(Q_\al^r)_+,Q_\beta^s]) \hbox{\qquad}  (\al<\beta)\\
	\Tr(-[(Q_\al^r)_-,Q_\beta^s]) \hbox{\qquad}(\al\geq\beta)\ea\right.$$

We take first the case $\al$=even, ($Q_\al\simeq
\em$) and $\beta$=odd ($Q_\beta\simeq I_+$) and the two point function
is written as:
			\a
	<\tau_{\al,r}\tau_{\beta,s}>=\Tr([Q_\al^r,Q_\beta^s]) \ \ \ \
	(\al>\beta \hbox{\ and zero otherwise})  \label{twopointtrace}
			\b

Here the number of $Q_\al$ and $Q_\beta$ in each trace must be the same, to
balance the $I_+$ and $\epsilon_-$'s (remember that $\al$ and $\beta$ have
different parity) so that $r$ and $s$ are forced to be equal.
The traces can be evaluated as above and one gets:
	\a
	<\tau_{\al,r}\tau_{\beta,s}>&=&
	(q_\al q_\beta)^r \Tr[(\em)^r,I_+^r]\delta_{r=s}=\0\\
	&=&(q_\al q_\beta)^r \delta_{r=s}
	    \left(\sum_{n=1}^{N-r}-\sum_{n=1}^N\right) n(n+1)...(n+r-1)
	\b
where $q_\al= {c_2\over c_1}{c_4\over c_3}...{1\over c_{\al-1}}$,
     $q_\beta={c_1\over c_2}{c_3\over c_4}...{c_{\beta-2}\over c_{\beta-1}}$.

When $\al$ is odd and $\beta$ even the result is the same with $r$ exchanged
with $s$. When $\al$ and $\beta$ are either both even or both odd
the 2-point function identically vanishes.

As an example:

$<\tau_{1,r}\tau_{2,s}>= \delta_{rs} \left(\sum_{n=1}^{N-r}-
\sum_{n=1}^N\right) n(n+1)...(n+r-1)$

$<\tau_{5,r}\tau_{2,s}>=0$  because $\al=2<\beta=5$.

\subsubsection{Quadratic models}

The 2-point correlation functions is  a  very  important  quantity  in
matrix models because  its  singularity  indicate  the  existence  of
critical points and its scaling near  them evaluate  the  anomalous
dimensions of corresponding operators.
In our case 2-point correlation  functions  permits  also  direct  the
calculation of the metric for the associated topological  model  (when
the puncture operator is $Q={\d \over \d N}$).

Using again the equation (\ref{2p}) we write down the two-point
correlation functions:
\a
<\tau_{1,r}\tau_{\al,s}>=\mbox{Tr}[(Q_1^r)_+,(Q_{\al}^s)_-]
\b

Using the form of the $Q$ matrices (\ref{Qalpha},\ref{sollinsys}) we
get the result:

\a
&&<\tau_{1,r}\tau_{\al,s}>=
\sum_{{\ba{c}0\leq 2l\leq r\\0\leq 2l'\leq s\ea}}
\sum_{{\ba{c}0\leq k\leq l\\0\leq k'\leq l'\ea}}
\sum_{{\ba{c}0\leq i\leq l-k\\l'-k'\leq j\leq2(l'-k')\\i+j
=l+l'-k-k'\ea}}\0\\
&&\left(\ba{c}r\\2l\ea \right)\left(\ba{c}s\\2l'\ea \right)
\left(\ba{c}2l-2k\\i\ea \right)\left(\ba{c}2l'-2k'\\j\ea \right)
(-1)^{k+k'}A_k^{(2l)}A_{k'}^{(2l')}f(S_1,T_1,S_{\al},T_{\al},R_{\al})\0\\
&&i!j!\sum_{n=0}^{N-1}\left[\left(\ba{c} n+2l-2k-i\\i\ea \right)\right.
\left(\ba{c}n+j\\j\ea \right)
-\left(\ba{c} n+2l'-2k'-j\\j\ea \right)\left.
\left(\ba{c}n+i\\i\ea \right)\right]\0
\b
where $f(\al)$ is:
\a
f(\al)=(g_1
h_\al/g_al)^{i+k}g_\al^{l'+l}s_0^{r-2l}h_\al^{l'-l}s_\al^{s-2l},\qquad
\al=1 \ldots q\0
\b

To calculate the 2-point correlation we needed the quantity
$\Tr  (I_+^n\epsilon_-^m I_+^p\epsilon_-^q)$. We use the fact that for $(n>m)$:
\a
I_+^n\epsilon_-^m=\sum_{k=0}^m \epsilon_-^k I_+^{n-m-k}A_k^{(n,m)},\quad
A_k^{(n,m)}={n!m!\over k!(n-m+k)!(m-k)!}
\b
Using this sum we get :
\a
\Tr  (I_+^n\epsilon_-^mI_+^p\epsilon_-^q)=
\sum_{k=0}^m{n!m!(q+k)!\over k!(n-m+k)!(m-k)!}\left(\ba{c}N\\ q+k+1\ea\right)
\delta_{n+p,m+q}
\b

The evaluation of higher genus contribution follows the same method  we
have followed at the calculation of 1-point correlation functions.
The only dependence of $N$ comes from
$\sum_{p=0}^{N-1}\left(\ba{c} p+s\\s\ea \right)
\left(\ba{c}p+n\\m\ea \right)$  and  we  are  looking  for  the
contribution of order $N^{1+k-2h}$.

We define the function $B(r,s|n,m)$ as the coefficient of the genus $h$
expansion:
\a
\sum_{p=0}^{N-1}\left(\ba{c} p+s\\s\ea \right)
\left(\ba{c}p+n\\m\ea \right)=N^{m+s-2h}B_h(r,s|n,m), \quad r+s=n+m
\b
The explicit expression is:

\a
&B_h(r,s|r',s')=\sum_{l=0}^{s+s'}\gamma_l(s,r,s')\left[{1\over 2}
\left(\ba{c} s+s'-l \\2h-l  \ea \right)(r-s)^{2h-l}+\right.\0\\
&+{(r-s)^{1+2h-l}\over s+s'-l+1}
\left(\ba{c} s+s'-l+1 \\ 1-2h-l \ea \right)+\0\\
&+\left.\sum_{2\leq 2t\leq s=s'-l}{B_{2t} \over 2t}
\left(\ba{c}s+s'-l\\2t-1\ea\right)
\left(\ba{c}s+s'-l-2t+1\\2h-l-2t+1\ea \right)(r-s)^{2h-l-2t+1}\right]\0
\b
$(r+r'=s+s')$; where $B_{2t}$ are Bernoulli numbers and $\gamma$ are:
\a
\gamma_l(s,r,s')=\sum_{k=0\ }^l
		\sum_{0\leq i_1\ldots i_k\leq s-1\ }
		\sum_{n-m\leq j_{k+1}\ldots j_l\leq n-1}
i_1\ldots i_k j_{k+1}\ldots j_l \0
\b
The genus $h$ contribution is:
\a
&&<\tau_{1,r}\tau_{\al,s}>_h=
\sum_{{\ba{c}0\leq 2l\leq r\\0\leq 2l'\leq s\ea}}
\sum_{{\ba{c}0\leq k\leq l\\0\leq k'\leq l'\ea}}
\sum_{{\ba{c}0\leq i\leq l-k\\l'-k'\leq j\leq2(l'-k')\\i+j
=l+l'-k-k'\ea}}\0\\
&&{r!s!(-1)^{k+k'}2^{-(k+k')}\over (r-2l)!(s-2l')!
(2l-2k-i)!(2l'-2k'-j)!k!k'!i!j!}f(\al)N^{l+l'-k-k'-2h}\0\\
&&(B_h(2(i+k-l)+j,j|2l-2k-i,i)-B_h(2(j+k'-l')+i,i|2l'-2k'-j,j))\0
\b

\section{Topological field theory properties of $2q$--matrix models}.

\setcounter{equation}{0}
\setcounter{subsection}{0}

We study in this section the content of $2q$--matrix models in terms
topological
field theories. The motivation is offered by the example of 2--matrix model,
which can be interpreted as a topological field theory with an infinite number
of primary fields, \cite{BX3}. We want to see whether a similar conclusion
can be drawn also for multi--matrix models.
The easiest way to identify a possible topological field theory (TFT) content
is to go to the cosmological point. We have seen previously that such a point
is not well defined for odd $q$ multi--matrix models. Consequently, in
this section, we concentrate on even $q$ multi--matrix models.
To be definite we
start with the 4--matrix model. We recall that the cosmological point is
identified by setting all the couplings to zero except the bilinear ones,
$c_{\al,\al+1}$, with reference to eq.(\ref{Z}). To simplify things further
we set from now on
\a
c_{\al,\al+1}= (-1)^\al\0
\b
without loss of generality (one can obtain the same results
by suitable rescaling the couplings of the discrete states). Finally we
replace $N$ by a continuous variable $x$ (i.e. we pass to a continuous
formalism
by suitably rescaling all the quantities and taking $N\to \infty$; $x$ is the
renormalized quantity that replaces $N$).

After these preliminaries let us concentrate on the 4--matrix models. Among
the discrete states, our candidates for primary states are $\{\psi_{a,b}, Q,
\omega_{c,d}\}$, where
\a
\psi_{a,b} = \chi_{a,0,b,0},\qquad \psi_{c,d}=\chi_{0,c,0,d}\0
\b
The relevant genus 0 correlators to study the TFT properties can be computed
from (\ref{2p}) and (\ref{3p})
\ai
<\psi_{a_1,b_1}\psi_{a_2,b_2}\omega_{c,d}> &=& \Big((a_1+b_1)(a_2+b_2)(c+d) -
c(a_1b_2+a_2b_1+b_1b_2) \0\\
&&-\frac{b_1b_2cd}{c+d-1}\Big)x^{c+d-1}\delta_{a_1+a_2+b_1+b_2,c+d}\label{3p1}\\
<\psi_{a,b}\omega_{c_1,d_1}\omega_{c_2,d_2}> &=& \Big((a+b)(c_1+d_1)(c_2+d_2)-
b(c_1d_2+c_2d_1+c_1c_2)\0\\
&&- \frac{abc_1c_2}{a+b-1}\Big)x^{a+b-1}
\delta_{a+b,c_1+c_2+d_1+d_2}\label{3p2}
\bj
and
\a
<Q\psi_{a,b}\omega_{c,d}> = (ac+ad+bd)x^{a+b-1}\delta_{a+b,c+d}\label{3pQ}
\b
We will also need $<QQQ>=x^{-1}$, which follows from the fact that
the correlators involving only $Q$ are the same as in the 2--matrix model,
see \cite{BX2}.

Now, at any level $r=a+b$ let us select an arbitrary state among the
$\psi_{a,b}$'s and call it $\psi_r, ~r>0$. Let us call ${\cal C}$ the
collection
of such choices for any $r$. Moreover, let us set $\omega_s\equiv
\omega_{0,s}$.
Then the states $\{\psi_r,Q,\omega_s\}$ constitute the set of primary states
of a TFT with puncture operator either $Q$ or $\psi_1$ or $\omega_1$.
This can be seen as follows. The non--vanishing structure constants are
\a
&&C_{r_1,r_2,\bar s}\equiv <\psi_{r_1}\psi_{r_2}\omega_s>= r_1r_2s
x^{s-1}\delta_{s,r_1+r_2}\0\\
&&C_{r,\bar s_1,\bar s_2}\equiv <\psi_r \omega_{s_1}\omega_{s_2}>=
rs_1s_2 x^{r-1}\delta_{r,s_1+s_2}\0\\
&&C_{0,r,\bar s}\equiv <Q\psi_r\omega_s> = rs x^{r-1}\delta_{r,s},
\qquad C_{0,0,0}\equiv<QQQ> =x^{-1}\0
\b
together with the ones obtained from these by permutation of the indices.
Now, if the puncture operator is $Q$, the metric is
\a
\eta_{r,\bar s}=\eta_{\bar s,r}
\equiv <Q\psi_r\omega_s>=rs x^{r-1}\delta_{r,s},\qquad \eta_{0,0}=
x^{-1}, \label{metric0}
\b
If the puncture operator is $\psi_1$, the metric is
\a
\eta_{r,\bar s}=\eta_{\bar s,r} \equiv <\psi_1\psi_r \omega_s>=
rs x^r \delta_{s,r+1},\qquad \eta_{0,\bar 1}=\eta_{\bar 1,0}=1 \label{metric1}
\b
The case when the puncture operator is $\omega_1$ is exactly specular to the
latter. These three cases, with exactly the same formulas for structure
constants and metric, were met in \cite{BX3}, where it was proven that the
inverse metric exists and the associativity conditions are satisfied.
The TFT obtained with a definite choice ${\cal C}$ will be denoted
${\cal T}_{\cal C}$. If necessary one can specify the symbol of the
relevant puncture operator.

Similarly, among the states $\omega_{c,d},~ c+d=s$ let us choose an
arbitrary one
and let us call it $\bar\omega_s,~s> 0$. Let us call $\bar{\cal C}$
such a choice for any level $s$. Moreover, let us set
$\psi_{r,0}\equiv \bar\psi_r$, $r> 0$. Once again
the states $\{\bar\psi_r,Q,\bar\omega_s\}$ constitute the primary states of
a TFT with puncture operator either $Q$ or $\psi_1$ or $\omega_1$.
We do not need to explicitly prove this since the formulas for the
structure constants and the metrics are the same as the previous ones
with the substitutions $\psi_r \rightarrow \bar\psi_r$ and
$\omega_s \rightarrow \bar \omega_s$. The TFT obtained with a definite
choice ${\bar{\cal C}}$ will be denoted ${\cal T}_{\bar{\cal C}}$.

We can think of ${\cal T}_{\cal C}$ and ${\cal T}_{\bar{\cal C}}$ as
unperturbed TFT's to which we couple
topological gravity. Therefore we are going to have puncture equations and
recursion relations. The latter are the same as in \cite{BX3} and will not be
repeated here. The former can be derived from the $W$ constraints. Instead
of writing the most general formula, we write down the simplest one for the
puncture operator $\psi_1$
\a
<\psi_1 \chi_{a_1,a_2,a_3,a_4}> = a_2<\chi_{a_1,a_2-1,a_3,a_4}>
+a_4 <\chi_{a_1,a_2,a_3,a_4-1}>\0
\b
from which one can infer the action of the puncture: $\psi_1$ lowers the
even indices by 1. Therefore, when $\psi_1$ is the puncture operator,
the descendants of $\psi_{a,b}$ are going to be $\chi_{a,n,b,m}$ for
positive $n$ and $m$, while any $\omega_{c,d}$ may be simultaneously primary
and descendant, or an isolated primary.

We notice that the situation here is an interesting generalization
of the situation in 2--matrix model, \cite{BX3}, where we have an infinity of
primary states denoted $\{T_n,Q,T_{-m}\}$, with nonnegative integer $n$ and
$m$,
where $T_n, T_{-m}$ are the discrete tachyonic states. Here we have $\infty^2$
primary states, which depend on two integral indices and could be referred
to as colored tachyons. In 2--matrix model, via reduction, one obtains
an infinite set of TFT models (the n--KdV models) coupled to topological
gravity, whose primary and descendants are to be found among the $T_n$'s
(or, symmetrically, among the $T_{-n}$'s), \cite{BCX}. Similarly here we expect
that, via reduction (see next section), the set $\psi_{a,b}$ with $a$ and $b$
positive, may support a series of matter TFT's coupled to topological gravity
(i.e. primaries and descendants). Due to its characteristics -- triangular
structure of the primaries and relation with the product of two n--KdV
models -- possible candidates (certainly not the only ones)
are, for example, the $W_3$ topological minimal models coupled
to topological gravity, \cite{Lerche},\cite{BLNW}.

In general, if we pass to $2q$--matrix models, the set of primaries will be
represented by the states $\{ \chi_{a_1,0,a_3,0,\ldots,a_{2q-1},0}\}$,
by $Q$ and by $\{\chi_{0,a_2,0,\ldots,0,a_{2q}}\}$:
the primaries are $\infty^q$.
This $q$ should be related to the target space dimension in a
string interpretation. In analogy with our previous discussion we are lead to
speculate that one of the two sets above can accomodate the states of
the $W_{q+1}$ minimal models coupled to topological gravity or anlogous TFT's.

\section{Non--Gaussian matrix models.}

The Gaussian version of $q$--matrix models is sufficient to study many
properties, in particular it is enough to identify the TFT character
of these models. From this point of view adding new interaction terms
amounts to switching on new perturbations, which is not a very interesting
complication in itself. However, if we come to reductions, i.e. if we try
to extract TFT's with a finite number of primaries from the infinite
TFT's that characterize the multi--matrix models, we are obliged to introduce
non--Gaussian interaction terms. In this paper we limit ourselves to
an example: our purpose is to show both the complexity inherent in
non--Gaussian perturbations and a possible way to circumvent it.

The example consists of switching on a cubic potential in the 4--matrix model.
More precisely we study the model ${\cal M}_{3,2,2,2}$.
The coupling conditions are the same as (\ref{4matrix}), except that in the
first equation a term $3v_1Q_1^2$ must be added, where $v_1\equiv t_{1,3}$
is the coupling of the cubic term in the potential. Consequently the coupling
conditions become a non--linear system of equations for the $Q_\al$'s.
Eliminating $Q_2$ and $Q_3$ we obtain:
\a
&&P_1+2(t_1-{c_1^2t_3\over 4t_2t_3-c_2^2})Q_1+
(u_1+c_1{2t_3u_2-u_3c_2\over 4t_2t_3-c_2^2}) -
{c_1c_2c_3\over 4t_2t_3-c_2^2}Q_4+ 3v_1 Q_1^2=0 \0\\
&&\overline{{\cal P}_4}+2(t_4-{c_3^2t_2\over 4t_2t_3-c_2^2})Q_4+
(u_4-c_3{2t_2u_3-u_2c_2\over 4t_2t_3-c_2^2}) -
{c_1c_2c_3\over 4t_2t_3-c_2^2}Q_1=0\label{m3222}
\b
We can therefore write this system in a simplified form as follows:
\a
&&P_1+3v_1 Q_1^2+2\tilde t_1Q_1+ \tilde u_1 + \tilde c Q_4+ =0 \0\\
&&\overline{{\cal P}_4}+2\tilde t_4Q_4+ \tilde u_4-\tilde c Q_1=0\label{m32}
\b
These are formally the coupling conditions of the model ${\cal M}_{3,2}$ with
$Q_2$ replaced by $Q_4$, the couplings being suitably renormalized, \cite{BCX}.

With reference to the coordinatization (\ref{jacobi}) the equations (\ref{m32})
in genus 0 become
\a
&&a_1(n) = - \frac {2\tilde t_4}{\~ c} R(n), \qquad
b_0(n) = - \frac{\~ u_4 + \~ c a_0(n)}{2 \~ t_4}\0\\
&& b_1(n) = - \frac{n + \~ c R(n) }{2 \~ t_4}, \qquad
b_2(n) = - \frac{3v_1}{\~ c} R(n)^2\0
\b
and the recursion  relations
\a
&& 2a_0(n) = -\frac{2\~ t_1}{3v_1} + \frac {\~ c}{6\~ t_4v_1}
\Big( \~ c + \frac{n}{R(n)}\Big)\label{recur1'}\\
&&2R(n) = \frac{\~ c}{2\~ t_4} a_0(n)^2 + \Big( \frac{2 \~ c \~ t_1}
{6\~ t_4 v_1} - \frac{\~ c^3}{12 \~ t_4^2 v_1}\Big) a_0(n)
-\frac{\~ c^2 \~ u_4}{12 \~ t_4^2v_1} + \frac {\~c \~ u_1}
{6\~ t_4v_1} \label{recur2'}
\b
We have therefore to solve a cubic equation. Once we have done this all the
unknowns can be determined and the explicit form of the matrices $Q_\al$ can
be calculated. The correlators (in genus 0) can be obtained as integrals
of algebraic equations. Writing down such expressions is not very interesting.
We can however ask ourselves whether in some region of the coupling space
we can find some interesting solution. This is actually the case.

Let us first simplify the formulas by imposing $6v_1=-1$, $\~ t_1=\~ u_4 =0$.
It can be shown, \cite{BCX}, that this is no loss of
generality. Then we impose the constraint
\a
a_0=0\label{constr1}
\b
The above equations then imply
\a
a_1 =\~ u_1 , \qquad R(n)= - \frac{n}{\~ c}, \qquad b_0 = b_1=0,
\qquad b_2(n) = \frac{n^2}{2\~ c^3}\0
\b
and, therefore,
\a
\~ u_1 = \frac{2\~ t_4}{\~ c^2} n \label{constr2}
\b
In general, when we impose constraints on the fields or on the coupling space,
we are not allowed to use the flow equations of the integrable hierarchy
(in this case the Toda lattice hierarchy), because such constraints might
deform the dynamics in a non--integrable manner. However one can prove that,
in the case of the constraint (\ref{constr1}) or, which is the same,
(\ref{constr2}), the constrained dynamics is still integrable and coincides
with the Toda flows constrained by eq.(\ref{constr1}): the resulting hierarchy
is the KdV hierarchy, \cite{BCX}. In other words, if we look at our 4--mm
in the submanifold of the parameter space specified by eq.(\ref{constr2}),
the correlators are those of the KdV model. As is well known this model
has only one primary field, $\psi_1$, and, in particular,
$<\psi_1\psi_1>=a_1=\~ u_1$.

By considering the model ${\cal M}_{p+2,2,2,2}$ one can
find the p-th critical point of the KdV series. One can also identify higher
KdV models. These are all (trivial) generalizations of the 2--mm. However, as
we
have remarked above, the structure of the q--matrix models allows for
more complex and interesting reductions which will be the object of future
research.

\section{The discretized 1D string}

\setcounter{equation}{0}
\setcounter{subsection}{0}

It is interesting to return to the problem of discretized
1D string theory within the present formalism and see how we recover
the results already obtained with other methods.

Let us start by studying and solving the following Gaussian partition function:
\a
Z=\int DM_i \exp\left[ Tr\left({\tilde{t}\over 2}\sum_{i=1}^{2q}M_i^2+c
\sum_{i=1}^{2q-1}M_iM_{i+1}\right)\right]\label{Zchain}
\b
The corresponding coupling conditions are ($t=\tilde{t}/c$):
\a
P_1 &+&c tQ_1+ c Q_2=0,\\ \0
tQ_{i}& +& Q_{i-1}+Q_{i+1}=0,
\ \ i=2,\ldots 2q-1\\
\overline{{\cal P}_{2q}} &+&c t Q_{2q}+c Q_{2q-1}=0 \0
\label{capcond}
\b
We express all $Q_{\al}$ matrices $\al=2,\ldots 2q-1$ in terms of $Q_1$ and
$Q_{2q}$. For this purpose we introduce the determinant of $n\times n$ matrix:
\a
D_n=\left|\ba{lccccr}
t&1&0&\ldots&\ldots&0\\
1&t&1&0&\ldots&0\\
0&1&t&\ddots&.&.\\
.&0&1&\ddots&1&0\\
0&\ldots&0&\ddots&t&1\\
0&&\ldots&0&1&t
\ea\right|\0
\b
The determinant $D_n$ satisfies the recursion relation:
\a
D_{n+1}=t D_n-D_{n-1}\0\\
D_0=1,D_1=t\0
\b
The expression of $D_n$ is :
\a
D_n=\sum_{k=0}^{[n/2]}t^{n-2k}(-1)^k \left(\ba{c}n-k\\k\ea\right)\0
\b
or:
\a
D_n={r_1^{n+1}-r_2^{n+1}\over r_1-r_2}\0
\b
where $r_1,r_2$ are the roots of the second order equation $r^2-tr+1=0$.

Solving the system  (\ref{capcond}) we get $(D_0=1)$:
\a
Q_i={(-1)^{i+1}\over D_{2q-2}}(Q_1D_{2q-i-1}-Q_{2q}D_{i-2}),i=2,\ldots 2q-1\0
\b
For $Q_1, Q_{2q}$ we get the usual 2-matrix model with quadratic potential
coupling conditions:
\a
P_1+\left(ct-{D_{2q-3}\over D_{2q-2}}\right)Q_1+{c\over D_{2q-2}}Q_{2q}=0\0\\
\overline{{\cal P}_{2q}}+\left(ct-{D_{2q-3}\over D_{2q-2}}\right)Q_{2q}+
{c\over D_{2q-2}}Q_{1}=0\0
\b
The  $Q_1, Q_{2q}$ matrices are:
\a
Q_1=I_++g_1(n)\em,\0\\
 Q_{2q}=h_{2q}(n)I_++R_{2q}(n)\em\0
\b
where:
\a
I_+=\sum_{n=0}^\infty E_{n,n+1},\em=\sum_{n=0}^\infty nE_{n,n-1},\0
\b
with
\a
g_1(n)=h_{2q}(n)R_{2q}(n)=-2{(ct D_{2q-2}-D_{2q-3})D_{2q-2}\over B},
\quad R_{2q}(n)=
{c D_{2q-2}\over B}\0
\b
and
\a
B=c^2(4t^2-1)D_{2q-2}^2-8ct D_{2q-2}D_{2q-3}+D_{2q-3}^2\0
\b
The form of $Q_\al$ matrices is:
\a
Q_\al=h_{\al}(n)  I_++g_{\al}(n)  \em,\ \quad \al=2\ldots 2q-1\0
\b
with\a
h_\al(n) ={(-1)^{\al+1}\over cB}[cD_{2q-\al-1}-
(ctD_{2q-2}-D_{2q-3})D_{\al-2}]\0\\
g_\al(n) =
{(-1)^{\al+1}\over B}[(ctD_{2q-2}-D_{2q-3}) D_{2q-\al-1}-cD_{\al-2}]\0
\b
Using (\ref{1p}), (\ref{2p}), we can now calculate the 1-point correlation
functions:
\a
<\tau_{\al,2r}>=\mbox{Tr}(Q_{\al}^{2r})=
\sum_{k=0}^r{(-1)^k2^{-k}(2r)!\over k!(r-k)!}
\left(\ba{c}N\\r-k-1\ea\right)(h_\al(n) g_\al(n))^r\0
\b
and 2-point correlation functions:
\a
<\tau_{1,2r}\tau_{\al,2s}>=\mbox{Tr}[(Q_1^{2r})_+,(Q_{\al}^{2s})_-]\0
\b
The 2-point correlation function is :
\a
<\tau_{1,2r}\tau_{\al,2s}>=
\sum_{\ba{c}0\leq k\leq r\\0\leq k'\leq s\ea}
\sum_{\ba{c}0\leq i\leq r-k\\s-k'\leq j\leq 2(s-k')\\i+j=r+s-k-k'\ea}
L_{rs}(k,k',i,j|N) f(\al)\0
\b
with:
\a
L_{r,s}(k,k',i,j|N)={(2r)!\over (r-2l)!(2l-2k-i)!k!}
{(2s)!\over (s-2l')!(2s-2k'-j)!k'!} (-1)^{k+k'}2^{-k-k'}\0\\
\sum_{n=0}^{N-1}\left[\left(\ba{c} n+2l-2k\\i\ea \right)
\left(\ba{c}n+2i+2j\\j\ea \right)-
\left(\ba{c} n+2l'-2k'\\j\ea \right)
\left(\ba{c}n+2i+2j\\i\ea \right)\right]\0
\b
and where $f(\al)$ is:
\a
f(\al)=
(g_1 h_\al/g_\al)^{i+k}(g_\al)^{r+s}
(h_\al)^{s-r},\al=2\ldots 2q-1\0
\b

We can obtain a more explicit formula for the determinant $D_n$ when
$2\geq t\geq -2$. In this case the roots of the equation $r^2-tr+1=0$ are
complex $r_{1,2}=\exp(\pm i \omega/2)$ and the
determinant is :
\a
D_n= {\sin(n\omega/2)\over sin(\omega/2)}\0
\b
with $\omega=(1/2)\mbox{arctan}\sqrt{(2/t)^2-1}$.
This formula permits us to single out the dependence of the 1-point
correlator
$<\tau_{\al,2r}>$ on the parameter $\al$. First we calculate:
\a
h_\al(n) g_\al(n)=(1/2)[(2A+A')+
(A'-2A\cos(\omega(q-1)))\cos(\omega(q-\al))]\0
\b
with
\a
A={D_{2q-3}-ctD_{2q-2}\over B},
A'={c^2D_{2q-2}^2-(ctD_{2q-2}-D_{2q-3})^2\over Bc}\0
\b
Hence we have the following behaviour:
\a
(h_\al(n) g_\al(n))^r=\sum_{k=-r}^r
d_k e^{i\omega k\al}\0
\b

Now we pass from the discrete variable $\al=1,\ldots 2q$ to a continuous time
$t\in [0,T/2]$. We introduce the puncture operator:
\a
O^{(2p)}=\int_0^{t/2}dt\mbox{Tr}(Q^{2r}(t))e^{ipt}=K_r
\sum_{k=-r}^r d_k \delta (p+\omega k)\0
\b
where $K_r$ behaves like $N^{r+1}$.

In the pure chain models (no potentials $V_{\al}$),
$Q(t)$ is proportional to either $I_+$ or $I_-$. From this it follows that
 $\mbox{Tr}(Q^r(t))$ is independent of $t$. Hence:
\a
O^{(2p)}\sim\delta (p)\0
\b
The conclusion is that for quadratic models we have, apart from the
fundamental state with zero momentum, also other excited states (discrete
states) with
integer momenta $p=n\omega,$ $n$ integer.

We can study the 2-point correlation functions in the same framework.
We take the particular case:
\a
<\tau_{1,2r}\tau_{\al,2r}>=\sum_{n=0}^r M_n (g_\al)^n
(h_\al)^{2r-n}\0
\b
with $n=i+k$ and:
\a
M_n=\sum_{0\leq k,k'\leq r }
L_{2r,2r}(k,k',n-k,2r-k'-n|N)(g_1)^n\0
\b
and look at the dependence on the parameter $\al$.
Using the dependence on $\al$ for :
\a
(g_\al)^n (h_\al)^{2r-n}=\sum_{k=-r}^{r}K_r^{(k)}
\exp(i\omega \al)\0
\b
we can calculate the 2-point correlation function.
Passing to the continuous time and using the symmetry of $K_r^{(k)}
K_{-r}^{(k)}$ we can write the expression of the 2-point correlation function,
as:
\a
<\tau_{1,2r}\tau_{\al,2r}>(\al)=\sum_{k=0}^r K_r^{(k)}\sin
\omega k(\al-\al_1)\0
\b
We can now evaluate 2-point correlators of puncture operators in the
momentum space:
\a
G^{(2r)}(p)=\int_0^{T/2}{dt\over 2\pi}e^{-p(t-t_1)}
<\tau_{1,2r}\tau_{\al,2r}>(t)=\sum_{k=0}^r {K_r^{(k)}\over p^2+
(k\omega)^2}
\label{gfunction}
\b
We have intermediate states at all integer momenta $p=k\omega$, $k$ integer.
The pulsation $\omega$ depends on the scaling we use when passing from
discrete values of $\al=1,\ldots 2q$ to
continuous time  $t\in [0,T/2]$ .

We can now apply all this to the $c=1$ string theory model model,
\cite{G}\cite{GKN}. The $c=1$ model with discrete time can be formulated
as a multi-matrix model with the partition function:
\a
Z=\int dM_i \exp\left[-{\beta\over 2}Tr\left(\sum_{i=1}^{n-1}
{(M_{i+1}-M_i)^2\over \epsilon}+\epsilon\sum_{i=1}^n V(M_i)\right)\right]\0
\b
with a quartic potential $V(M)=M^2-gM^4$. However, only the
contribution near saddle point $V'(M_c)=0$, where the potential is quadratic
in the fluctuation $\Delta M$, is essential
\a
V(M)={1\over 4g}-{2(\Delta M)^2 \over \beta}, M=M_c+
{\Delta M \over \sqrt{\beta}}
\b
The new partition function is (up to the constant $\exp(-N\beta\epsilon/(8g))$:
\a
Z=\int dM_i \exp\left[ Tr\left(\sum_{i=1}^n
\Delta M_i^2(2\epsilon-{1\over \epsilon})+
{1\over \epsilon}\sum_{i=1}^{n-1}\Delta M_i \Delta M_{i+1}\right)\right]\0
\b
It represents a string theory on circle with radius $R\sim {1\over \epsilon}$.
This is exactly our initial partition function (\ref{Zchain}) with the
identifications $\tilde{t}=2(2\epsilon-1/\epsilon)$, $c=
1/\epsilon$ than $t=2(2\epsilon^2-1)$ and we have the following limiting cases
for the determinant $D_n$:
\ai
D_n\sim t^n ~~~~~~&{\rm for}&\epsilon\rightarrow \infty,t\rightarrow \infty
\0\\
D_{3n+k}\sim (-1)^n(1+k),\quad k=0,-1,-2 &{\rm for}&
\epsilon\rightarrow 0,t\rightarrow 2(2\epsilon^2-1)\0
\b
{}From the results before (\ref{gfunction}) we have found that our model
describes particles with energy levels equal to
$ (n+1/2)\omega(\epsilon)/\beta$. For small $\epsilon$, $\omega\sim
\epsilon$ and changing $\epsilon$ (lattice spacing) means a linear change
of energy scale.
In this limit the model describes a string with the discretized
real line as target space.

Because $sin(\omega/2)=2\epsilon\sqrt{1-\epsilon^2}$, for $\epsilon\leq 1$
the pulsation becomes $\omega$ complex, which is a sign of instability
of the model. For
$\epsilon\rightarrow \infty$ the model decouples in $q$ noninteracting
1-matrix models.
The instability is due to the liberation of the vortices and give rise to
the Kosterlitz-Thouless transition for $\epsilon$ near 1.

Our method allows us to calculate (at least in principle) all $n$-point
functions at any genus and
what is more important it permits calculations in the vortex region where
$\epsilon\leq 1,~t\geq 2$.

\section*{}
\setcounter{equation}{0}
\setcounter{subsection}{0}

\subsection*{Appendix A. The W--constraints}

This Appendix is devoted to the derivation of the W--constraints in
$q$--matrix models.
{}From both the coupling equations (\ref{coup3}) and consistency conditions
(2.15), we get
the W-constraints in the form:
$Tr(Q^{n+r}(\alpha)\partial_{\lambda_{\alpha}}^r(*))=0$
where $*$ are the relations (2.9) \footnote{For another approach see
\cite{IM}.}.

W-constraints have the form:
\a
W_n^{[r]}(\alpha)Z_N(t,c)=0,\ r\geq 0,n\geq -r;\quad\alpha=1,\ldots q.
\b
or
\a
({\cal L }_n^{[r]} (\alpha)-(-1)^r T_n^{[r]}(\alpha))Z_N(t,c)=0.\0
\b
involving the interaction operator  $T_n^{[r]}$ which depends only on all the
couplings $g_{a_1\ldots a_q}$, except $g_{0,...,0,a_\al,0,...,0}=
t_{\al,a_\al}$.

For example $T_n^{[1]}$ and $T_n^{[2]}$  are:
\a
T_n^{[1]}(\alpha)&=&a_{\alpha} g_{a_1\ldots a_q}
\frac {\partial}{ \partial g_{a_1\ldots,a_{\alpha}+n,\ldots a_q}}\\
T_n^{[2]}(\alpha)&=&a_{\alpha}a_\al' g_{a_1\ldots a_q} g_{a_1'\ldots a_q'}
\frac {\partial}{ \partial g_{a_1+a_1'\ldots,a_{\alpha}+a_\al'+n,\ldots a_q+
a_q'}} +\0\\
&+&a_{\alpha}(a_\al-1) g_{a_1\ldots a_q}
\frac {\partial}{ \partial g_{a_1\ldots,a_{\alpha}+n,\ldots a_q}}\0
\b
The operator ${\cal L}_n^{[r]}(1) $ has the same form as that of the two-matrix
model:
\a
{\cal L}_n^{[r]}(1)=\int dz:\frac {1}{ r+1}(\partial_z+J)^{r+1}:z^{r+n}
\b
where $::$ is the normal ordering and $J(z)$ is the $U(1)$ current:
\a
J(z)=\sum_{k=1}^{p_1} k t_{1,k} z^{k-1} +N z^{-1 }+\sum_{k=1}^{\infty} z^{-k-1}
\frac{\partial }{\partial t_{1,k}}
\b
The same expression holds for ${\cal L}_n^{[r]}(q)$.

The expression of ${\cal L}_n^{[r]}(\alpha),\alpha=2,\ldots q-1$ is different
due to the absence of the $P$-matrix term:
\a
{\cal L }_n^{[r]}(\alpha)=\int dz:\frac {1}{ r}(\partial_z+V'_{\alpha})^r
 P_{\alpha}:z^{r+n }
\b
with
\a
V'_{\alpha}=\sum_{k=1}^{p_{\alpha}} k t_{\alpha,k }z^{k-1} +N z^{-1},\\ \0
P_{\alpha}=N z^{-1} +\sum_{k=1}^{\infty}z^{-k-1}
\frac{\partial }{\partial t_{\alpha,k }}\0
\b
The explicit expression of the first terms is:
\a
{\cal L }_n^{[1]}(\alpha)&=&\sum_{k=1}^\infty kt_{\al,k}{\d\over
 \d t_{\al,k+n}}+
Nt_{\al,1}\delta_{n,-1}\0\\
{\cal L }_n^{[2]}(\alpha)&=&\sum_{k=1}^\infty k(k-1)t_{\al,k}{\d\over
\d t_{\al,k+n}}+
\sum_{k_1,k_2} k_1k_2t_{\al,k_1}t_{\al,k_2}{\d\over \d t_{\al,k+n}}+\0\\
&+&N^2t_{\al,1}\delta_{n,-1}+N(t_{\al,1}^2+2t_{\al,2})\delta_{n,-2}\0
\b

As an example we write down the $W^{[1]}_{-1},W^{[1]}_0$ and $W^{[1]}_1$
constraints for the three matrix model.

$W^{[1]}$:
\def \l{\left<}
\def \r{\right>}
\a
\sum kt_k\l\tau_{k-1}\r+Nt_1+c_{12}\l\lambda_1\r+c_{13}\l\sigma_1\r&=&0\0\\
\sum ku_k\l\lambda_{k-1}\r+Nu_1+c_{12}\l\tau_1\r+c_{23}\l\sigma_1\r&=&0\\
\sum ks_k\l\sigma_{k-1}\r+Ns_1+c_{23}\l\lambda_1\r+c_{13}\l\tau_1\r&=&0\0\\
\0\\
\sum kt_k\l\tau_k\r+c_{12}\l\chi_{110}\r+c_{13}\l\chi_{101}\r&=&
-{N(N+1)\over2}\0\\
\sum ku_k\l\lambda_k\r+c_{12}\l\chi_{110}\r+c_{23}\l\chi_{011}\r&=&0\\
\sum kt_k\l\sigma_k\r+c_{13}\l\chi_{101}\r+c_{23}\l\chi_{011}\r&=&
-{N(N+1)\over2}\0\\
\0\\
\sum kt_k\l\tau_{k+1}\r+(N+1)\l\tau_1\r+c_{12}\l\chi_{210}\r+
c_{13}\l\chi_{201}\r&=&0\0\\
\sum ku_k\l\lambda_{k+1}\r\ \ \ \ \ \ \ \ \ \ \ \ \ +
c_{12}\l\chi_{120}\r+c_{23}\l\chi_{021}\r&=&0\\
\sum kt_k\l\sigma_{k+1}\r+(N+1)\l\sigma_1\r+c_{13}\l\chi_{102}\r+
c_{23}\l\chi_{021}\r&=&0\0
\b

One easily sees from the second group of identities that the limit of
pure chain models (cosmological point) does not exists for three--mm.
The same thing holds for odd--q matrix models. However, writing down the
W constraints for even--q matrix models, one can see that such a limit exists.
This confirms the results obtained with other methods.

\subsection*{Appendix B. Explicit derivation of 1p correlators.}

In this Appendix
we give the derivation of the 1--point functions promised in section 4.
For this we need to know $Q^p$
where $Q=I_++a \em$;we express it in terms of the normal ordered
quantities $:Q^l:$ ($QQ=:QQ:+[QQ]$)
\a
Q^p=\sum_{k=0}^{[p/2]}:Q^{p-2k}:([QQ])^k A_k^{(p)}
\b
where the contractions are $[QQ]=-aI_0$ because $[I_+,\em]=I_0$ ;the normal
ordering means that we have expressions with $I_+$ on left side and $\em$
on the right side.

The
coefficient $A_k^{(p)}$ is the number of ways in which we can choose k pairs
from p identical objects:
\a
A_k^{(p)}={1\over k!}\left(\ba{c}p\\2\ea \right)\left(\ba{c}p-2\\2\ea \right)
\ldots \left(\ba{c}p-2k+2\\2\ea \right)={p!2^{-k}\over (p-2k)!k!}\0
\b
Using :
\a
:Q^{p-2k}:=\sum_{i=0}^{p-2k}\left(\ba{c}p-2k\\i\ea \right)a^iI_+^{p-2k-i}
(\em)^i
\b
We have the result:
\a
Q^{2p}=\sum_{k=0}^{2p}\sum_{i=0}^{2p-2k}\left(\ba{c}2p-2k\\i\ea \right)
(-1)^k a^{i+k} A_k^{(2p)}I_+^{2p-2k-i}(\em)^i
\label{qq}
\b
We define the $Q_1$ matrix:
\a
Q_1=I_++a_0I_0+a_1(\em)=Q(a_1)+a_0I_0 \0\\
Q_1^r=\sum_{2l=0}^r\left(\ba{c}r\\2l\ea \right)Q(a_1)^{2l}a_0^{r-2l}
\b
Using (\ref{qq}) the expression of $Q_1^r$ is:
\a
Q_1^{r}=\sum_{2l=0}^r\sum_{k=0}^{2l}\sum_{i=0}^{2l-2k}
\left(\ba{c}r\\2l\ea \right)\left(\ba{c}2l-2k\\i\ea \right)
(-1)^k  A_k^{(2l)} a_1^{i+k}a_0^{r-2l}I_+^{2l-2k-i}(\em)^i
\b
The same expressions are for :
\a
Q_{\al}^{r}=\sum_{2l=0}^r\sum_{k=0}^{2l}\sum_{i=0}^{2l-2k}
\left(\ba{c}r\\2l\ea \right)\left(\ba{c}2l-2k\\i\ea \right)
(-1)^k  A_k^{(2l)} f_1(\al)I_+^{2l-2k-i}(\em)^i\\
f_1(\al)=(g_\al/h_\al)^{i+k}h_\al^{2l}s_\al^{r-2l}, \al=1 \ldots q\0
\b
Because we have the summation:
\a
\mbox{Tr}(I_+^k(\em)^k)=k!\sum_{N=0}^{N-1}
\left(\ba{c}n+k\\k\ea \right)
=k!\left(\ba{c}N+k\\k+1\ea \right)\0
\b
the 1-point correlation function is:
\a
<\tau_r>=\mbox{Tr}Q_1^r=\sum_{2l=0}^{r}\sum_{k=0}^{l}
{(-1)^k 2^{-k}r!\over (r-2l)!k!(l-k)!}\left(\ba{c}N+l-k\\l-k+1\ea \right)
(h_\al g_\al)^l s_\al^{r-2l}
\b

\end{document}